\documentclass[
twocolumn,
letter]{jpsj3}
\usepackage{txfonts}
\usepackage{cite,overcite}
\usepackage{amsmath,amssymb,bm
}
\usepackage[dvipdfmx]{graphicx}
\usepackage{latexsym}
\usepackage{braket}
\usepackage{mathrsfs}
\usepackage{cancel}
\usepackage{color}
\usepackage{multirow}
\usepackage{mathtools}
\usepackage{mathrsfs}
\usepackage{comment}
\usepackage{cancel}
\renewcommand{\d}{{\rm d}}
\renewcommand{\i}{{\rm i}}
\newcommand{\e}{{\rm e}}

\usepackage{xcolor}

\usepackage{appendix}

\title{Lieb-Schultz-Mattis-Type and Laughlin-Type Argument for the
Quantum Hall Effect in Lattice Fermions with Spiral Boundary Conditions}

\author{Masaaki Nakamura$^{1*}$ and Masanori Yamanaka$^{2\dag}$}

\inst{
$^1$Department of Physics, Ehime University, Bunkyo-cho 2-5, Matsuyama,
Ehime 790-8577, Japan\\
$^2$Department of Physics, College of Science and Technology, Nihon
University, Kanda-Surugadai 1-8, Chiyoda-ku, Tokyo 101-8308, Japan
}

\abst{
We derive the condition for the occurrence of the integer quantum Hall
effect in two-dimensional lattice systems with interactions, expressed
as $\phi\nu-\rho\in\mathbb{Z}$, where $\phi$, $\nu$, and $\rho$ denote
the magnetic flux, the Chern number, and the electron density,
respectively.  By employing spiral boundary conditions, which treat the
system as an extended one-dimensional chain, this condition is obtained
directly through a Lieb-Schultz-Mattis-type and Laughlin-type argument.
This approach improves upon the preceding work based on conventional
periodic boundary conditions, where the condition was derived indirectly
with redundant system-size dependence.  The key to this approach is that
the spatial directions of the external force and the response can be
systematically controlled by a factor of the system size.
}

\begin{document}

\maketitle
{\it --Introduction--}

The quantization of the Hall conductance in two-dimensional (2D)
electron systems, known as the quantum Hall effect (QHE), is one of the
most remarkable manifestations of topology in condensed matter physics.
The quantized Hall conductivity is expressed in terms of the Chern
number of the occupied bands\cite{TKNN,Kohmoto1985,Niu-Thouless-Wu}.
From the viewpoint of many-body theory, the quantization of the Hall
conductance can also be derived from Laughlin's gauge
argument,\cite{Laughlin} which explains the integer change of
polarization upon adiabatic flux insertion.
In lattice systems, the condition for the appearance of the QHE is
given by a Diophantine equation that relates the magnetic flux, the
Chern number, and the electron density.\cite{TKNN,Dana-Avron-Zak}

On the other hand, a similar topological constraint was discussed by
Lieb, Schultz, and Mattis (LSM) for one-dimensional (1D) quantum spin
systems,\cite{Lieb-S-M1961} and later generalized to various correlated
systems\cite{Affleck-L,Affleck,Oshikawa-Y-A,Yamanaka-O-A} and higher
dimensions\cite{Oshikawa-2000a,Oshikawa-2000b,Oshikawa-2003,
Hastings2005,Hastings2004,Yao-O}.  These approaches are closely related
through the concept of the modern theory of polarization introduced by
Resta\cite{Resta1994,Resta1998,Resta-S1999}.  The theory of polarization
has also been applied to various
systems\cite{Nakamura-V2002,Nakamura-T2002}, and also to the calculation
of topological numbers of 2D lattice systems by applying Laughlin's
arguments of flux insertion\cite{Coh-V}.

Lu, Ran, and Oshikawa derived the condition for the appearance of the
QHE in lattice systems with interactions by combining the LSM-type
argument with Laughlin-type argument as\cite{Lu-R-O,Matsugatani-I-S-W}
\begin{equation}
\phi\nu - \rho \in \mathbb{Z},
 \label{theorem1}
\end{equation}
where $\phi$, $\nu$, and $\rho$ denote the magnetic flux, the Chern
number, and the electron density, respectively.  In their theory based
on periodic boundary conditions (PBCs), the condition for QHE appears
indirectly passing through the intermediate relation which depends on an
artificial geometric parameter.  Therefore, it is desirable to improve
the proof to eliminate this redundancy.

\begin{figure}[t]
\begin{center}
 \includegraphics[width=6cm]{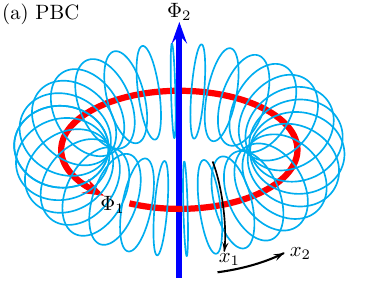}
 \includegraphics[width=6cm]{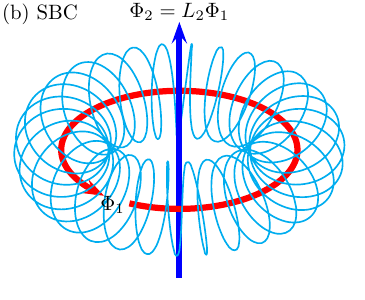}
\end{center}
 \caption{(Color) Boundary conditions for 2D lattices: (a) Conventional
periodic boundary conditions (PBCs), (b) Spiral boundary conditions
(SBCs), where the system is described as an extended 1D chain.  Fluxes
$\Phi_i$ for 2D lattices are applied as $\mathcal{H}_i(\Phi)=
U_i^{\Phi/2\pi} \mathcal{H} U_i^{-\Phi/2\pi}$. For SBCs, the insertion
of flux $\mathcal{H}_1'(\Phi_1)$ has the same effect as the flux
$\mathcal{H}_2'(L_2\Phi_1)$ due to the relation $U_1'=U_2^{\prime L_2}$.
}
%
\label{fig:torus}
\end{figure}

In this Letter, we reformulate the argument using \textit{spiral
boundary conditions}
(SBCs)\cite{Yao-O,Nakamura-M-N,Nakamura-M,Kadosawa-N-O-N1}, which allow
a 2D system to be regarded as an extended 1D chain (see
Fig.~\ref{fig:torus}).  By employing the LSM-type and Laughlin-type
arguments under the SBC, we derive the condition for QHE
(\ref{theorem1}) \textit{directly}, without invoking any system-size
dependence.  This formalism provides a unified and more transparent
understanding of the QHE from the standpoint of symmetry and topology.

In the rest of this Letter, we explain the argument by
Ref.~\cite{Lu-R-O}. Then we reformulate this argument based on SBCs and
derive Eq.~(\ref{theorem1}).  Throughout this Letter, the electron
charge, the reduced Planck constant, and the lattice constant are set to
be unity as $e=\hbar=a=1$.


{\it --Conventional periodic boundary conditions--}

\begin{figure}[t]
\begin{center}
 \includegraphics[width=9cm]{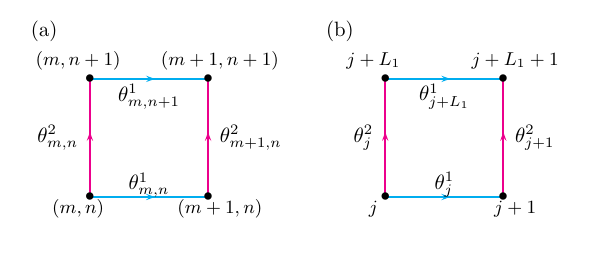}
\end{center}
 \caption{(Color) The gauge field $\theta^{\alpha}$ in a square lattice
in a magnetic field with (a) periodic boundary conditions, and (b)
spiral boundary conditions.}
\label{fig:plaquette}
\end{figure}

First, we discuss how the condition given by Eq.~(\ref{theorem1}) is
derived following the discussion by Lu, Ran and Oshikawa\cite{Lu-R-O}.
The Hamiltonian for the fermions in the 2D square lattice with
$L_1\times L_2$ sites under periodic boundary conditions is given
by\cite{Hofstadter}
\begin{equation}
 \mathcal{H} =-t \sum_{m,n}
  (c^\dag_{m+1,n}\e^{\i\theta_{m,n}^1}c^{\mathstrut}_{m,n}
  +c^\dag_{m,n+1}\e^{\i\theta_{m,n}^2}c^{\mathstrut}_{m,n}
  +\mbox{H.c.})+V(\{n_{m,n}\}),
  \label{Ham_PBC}
\end{equation}
where $c^{\mathstrut}_{m,n}$ ($c_{m,n}^{\dag}$) is the annihilation
(creation) operator at $(m,n)$ site, $t$ is the hopping amplitude,
$V(\{n_{m,n}\})$ represents the fermion interaction and is written in
terms of the number operator $n_{m,n}\equiv
c_{m,n}^{\dag}c_{m,n}^{\mathstrut}$. The gauge field is introduced in
the Landau gauge as
\begin{equation}
 \theta^1_{m,n}=0,\qquad \theta^2_{m,n}=2\pi\phi m,
\end{equation}
which gives the magnetic flux $B=-\frac{2\pi\hbar}{ea^2}\phi$
with
\begin{equation}
 \phi=\frac{1}{2\pi}\sum_{\Box}\theta_{m,n}
  =\frac{p}{q},
  \label{box}
\end{equation}
where $p$ and $q$ are coprime integers.  The summation for the plaquette
is taken as shown in Fig.~\ref{fig:plaquette}(a),
\begin{equation}
 \sum_{\Box}\theta_{m,n}
  =\theta_{m,n}^1+\theta_{m+1,n}^2-\theta_{m,n+1}^1-\theta_{m,n}^2.
\end{equation}

\begin{figure}[t]
\begin{center}
 \includegraphics[width=3.8cm]{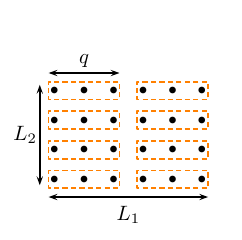}
\end{center}
 \caption{(Color) Sublattice structure of the system
 $(L_1,L_2,q)=(6,4,3)$ with a magnetic field ($\phi=p/q$). Here, $L_1/q$
 is chosen to be an integer.}
\label{fig:sublattice}
\end{figure}

In the absence of a magnetic field ($\phi=0$), the Hamiltonian possesses
translational symmetry by one lattice spacing along both directions,
$\mathcal{H}=T_i^{\mathstrut}\mathcal{H}T_i^{-1}$. For $\phi\neq 0$,
this translational symmetry is broken due to the position dependence of
$\theta^2_{m,n}$ and sublattice structure appears as shown in
Fig.~\ref{fig:sublattice}. Therefore, $L_1$ is selected so that
$L_1/q\in\mathbb{Z}$. On the other hand, a similar symmetry is satisfied
as $\mathcal{H}=\tilde{T}_i^{\mathstrut}\mathcal{H}\tilde{T}_i^{-1}$ by
introducing the magnetic translation operators
\begin{equation}
 \tilde{T}_1=U_2^{L_2\phi}T_1,\qquad \tilde{T}_2=T_2.
  \label{mTrans}
\end{equation}
The polarization operator
\begin{equation}
 U_i=\exp\left[\i\frac{2\pi}{L_i}\sum_{\bm{x}}x_i n_{\bm{x}}\right].
  \label{Ui}
\end{equation}
with $\bm{x}=(x_1,x_2)$ satisfies the relation
\begin{equation}
 T_1 U_1 T_1^{-1}
  =\e^{-\i 2\pi L_2\hat{\rho}}U_1.
\label{TUT.1}
\end{equation}
where $\hat{\rho}$ is the density operator, and $L_2$ represents the
{\it cross section} of the 2D system. As in the LSM argument for $d\geq
2$, $L_2$ and $q$ should be chosen to be coprime.

Now we consider the quantum Hall effect in this lattice system.  The
quantized Hall conductivity is characterized by the Chern number $\nu$
and the density $\rho$ as
\begin{equation}
 \sigma_{xy}=\frac{e^2}{h}\nu,\qquad \rho=\frac{r}{q},
\end{equation}
where $r$ is an integer.  Since the magnetic flux $\phi=p/q$ induces a
sublattice structure with $q$ sites as shown in
Fig.~\ref{fig:sublattice}, a commensurate filling is required to open an
energy gap.  Thus, for the QHE where $\nu$ is an integer, the following
condition must hold:
\begin{equation}
 \phi\nu-\rho=n,\qquad n\in\mathbb{Z},
  \label{condition_QHE.1}
\end{equation}
which is equivalent to the Diophantine equation
\begin{equation}
 r=\nu p+nq.
\end{equation}
The Chern number $\nu$ can be obtained from Resta's formulation of
polarization with flux insertion to the ground state \cite{Coh-V,Lu-R-O}
$\ket{\Psi_i(0)}$ as shown in Fig.~\ref{fig:torus}(a):
\begin{equation}
 z(\Phi)=z(0)\e^{-\i \nu\Phi},\quad 
 z(\Phi_2)=\braket{\Psi_2(\Phi_2)|U_1|\Psi_2(\Phi_2)},
 \label{z_def}
\end{equation}
leading to
\begin{equation}
 \nu=-\frac{1}{\Phi}\mathrm{Im\,ln\,}z(\Phi).
\end{equation}
This is also an expression of Laughlin's argument for QHE
\cite{Laughlin} in lattice systems. It states that inserting a unit flux
$\Phi=2\pi h/e$ in the $x_2$ direction pumps an integer number of
fermions in the $x_1$ direction when $\sigma_{xy}$ is quantized.

The adiabatic flux insertion is expressed as
\begin{equation}
 \ket{\Psi_i(\Phi_i)}
  =F_i(\Phi_i)\ket{\Psi_i(0)},
  \label{FIa.1}
\end{equation}
with
\begin{equation}
  F_i(\Phi_i)
 =\mathcal{T}
  \exp\left[-\i\int_0^T\mathcal{H}_i(\Phi_i\tfrac{t}{T})\d t\right],
  \label{FIb.1}
\end{equation}
where $\mathcal{T}$ denotes time ordering, $T$ is a sufficiently long
period compared with the inverse of the energy gap, and
\begin{equation}
 \mathcal{H}_i(\Phi)\equiv U_i^{\Phi/2\pi} \mathcal{H}
  U_i^{-\Phi/2\pi}.
  \label{fluxes_PBC}
\end{equation}
Equation (\ref{z_def}) then yields the operator relation
\begin{equation}
 [F_2(\Phi)]^{-1}U_1F_2(\Phi)=U_1 \e^{-\i \nu\Phi}.
  \label{FUF.1}
\end{equation}
Since $F_i(2\pi)$
plays a similar role to $U_i$, we can introduce a magnetic translation
operator as an analogy of Eq.~(\ref{mTrans}),
\begin{equation}
 \bar{T}_1=F_2(2\pi L_2\phi)T_1,
  \label{FT.1}
\end{equation}
which satisfies
$\mathcal{H}=\bar{T}_i^{\mathstrut}\mathcal{H}\bar{T}_i^{-1}$ for
$\phi\neq 0$.  We then obtain
\begin{equation}
 \bar{T}_1 U_1 \bar{T}_1^{-1}
 =U_1\e^{\i 2\pi L_2(\phi\nu-\rho)}.
 \label{TUT.1a}
\end{equation}
Thus, following an LSM-type argument for the symmetry operation, the
condition for $z(\Phi)\neq 0$ becomes
\begin{align}
 z(\Phi)
 &=\braket{\Psi_2(\Phi_2)|\bar{T}_1U_1\bar{T}_1^{-1}|\Psi_2(\Phi_2)}\\
 &=z(\Phi)\e^{\i 2\pi L_2(\phi\nu-\rho)}.
 \label{LSM_z}
\end{align}
Hence,
\begin{equation}
 L_2(\phi\nu-\rho)\in\mathbb{Z},
  \label{LSM_condition1}
\end{equation}
Since this must hold for arbitrary integer $L_2$, we finally obtain the
condition $\phi\nu-\rho\in\mathbb{Z}$ which is precisely the condition
for the QHE given by Eq.~(\ref{theorem1}). However, the appearance of
the artificial parameter $L_2$ is undesirable, because $L_2$ must be
chosen to be relatively prime to $q$ so that a redundancy remains.


{\it --Spiral boundary conditions--}

Next, we reconsider the above discussion under spiral boundary
conditions (SBCs), where the Hamiltonian (\ref{Ham_PBC}) is rewritten as
a 1D model with long-range hopping,
\begin{equation}
  \mathcal{H}' =-t \sum_{j=1}^{L}
  (c^\dag_{j+1}\e^{\i\theta_{j}^1}c^{\mathstrut}_{j}
  +c^\dag_{j+L_1}\e^{\i\theta_{j}^2}c^{\mathstrut}_{j}
  +\mbox{H.c.})+V(\{n_{j}\}),
\end{equation}
where $c^{\mathstrut}_{j}$ ($c_{j}^{\dag}$) is the annihilation
(creation) operator at $j$th site of the extended 1D chain of SBCs, and
$L\equiv L_1L_2$ is the total number of lattice sites.  The gauge field
is given by (see Fig.~\ref{fig:plaquette}(b))
\begin{equation}
 \theta^1_{j}=0,\qquad
 \theta^2_{j}=2\pi\phi j.
\end{equation}

\begin{figure}[t]
\begin{center}
 \includegraphics[width=6cm]{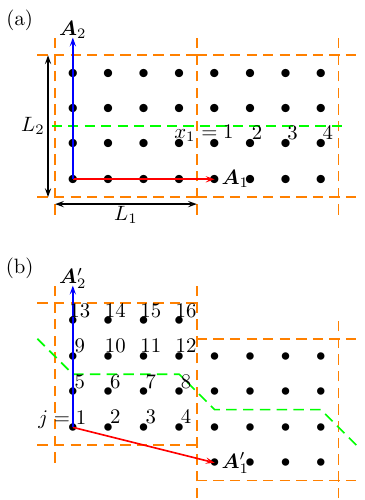}
\end{center}
 \caption{(Color) Boundary conditions for 2D square lattices with
 $L_1\times L_2$ sites: (a) Periodic boundary conditions (PBCs), (b)
 Spiral boundary conditions (SBCs).  For the systems with SBCs, the
 lattices are labeled as extended 1D chains.  $\bm{A}_i$ and $\bm{A}_i'$
 are primitive vectors.  When we cut the system (by green lines), the
 difference in the number of removed bonds in both boundary conditions
 is only one.}
\label{fig:unit_SBC}
\end{figure}

Now we introduce the polarization operators in SBCs.  For this purpose, we
reinterpret the polarization operators in PBCs (\ref{Ui}). The primitive
vectors in the real space (see Fig.~\ref{fig:unit_SBC}(a)) and the
reciprocal lattice space of 2D systems as
\begin{subequations}
\begin{alignat}{3}
 \bm{A}_1&=(L_1,0), &\qquad \bm{B}_1&=2\pi(\tfrac{1}{L_1},0),
 \label{AB1}\\
 \bm{A}_2&=(0,L_2), &  \bm{B}_2&=2\pi(0,\tfrac{1}{L_2}).
 \label{AB2}
\end{alignat}
\end{subequations}
Then these vectors satisfy the following relation,
\begin{equation}
 \bm{A}_{i}\cdot\bm{B}_{j}=2\pi\delta_{ij}.
  \label{defAB}
\end{equation}
The polarization operators (\ref{Ui}) are written using the above primitive
vectors in the reciprocal lattice space as
\begin{equation}
 U_{i}
  =\exp\left[\i\sum_{\bm{x}}\bm{B}_{i}\cdot\bm{x}n_{\bm{x}}\right].
  \label{Ui_gen}
\end{equation}
For SBCs as shown in Fig.~\ref{fig:unit_SBC}(b), the primitive vectors
are chosen so that they satisfy the relation (\ref{defAB}),
\begin{subequations}
 \label{A'B'}
\begin{alignat}{3}
 \bm{A}'_1&=(L_1,-1),&\qquad \bm{B}'_1&=2\pi(\tfrac{1}{L_1},0),
  \label{AB1'}\\
 \bm{A}'_2&=(0,L_2),&
 \bm{B}'_2&=2\pi(\tfrac{1}{L_1L_2},\tfrac{1}{L_2}).
 \label{AB2'}
\end{alignat}
\end{subequations}
The polarization operators in 2D systems with SBCs are introduced as
\begin{equation}
 U'_i
  =\exp\left[\i\sum_{\bm{x}}\bm{B}'_i\cdot\bm{x}n_{\bm{x}}\right].
  \label{LSMYO.22}
\end{equation}
For $i=1$, due to the relation $\bm{B}_1'=\bm{B}_1$, $U_1'$ is the same
operator as $U_1$. The representation in SBCs is given by the relation
\begin{equation}
 \frac{2\pi jn_{j}}{L_1}=\frac{2\pi x_1n_{\bm{x}}}{L_1}
  \quad (\mathrm{mod}\,2\pi).
  \label{U1mod}
\end{equation}
For $i=2$, the exponent part of the polarization operator can be written
in the SBC representation as
\begin{align}
 \sum_{\bm{x}}\bm{B}'_2\cdot\bm{x}n_{\bm{x}}
  &=\sum_{m=1}^{L_1}\sum_{n=0}^{L_2-1}
  2\pi\frac{m+nL_1}{L_1L_2}n_{m,n}\\
  &=\sum_{j=1}^{L}\frac{2\pi}{L}jn_{j},
\end{align}
where $j=m+nL_1$.  Thus the polarization operators for SBCs are identified as
\begin{equation}
 U_1'=\exp\left[\i\sum_{j=1}^L\frac{2\pi jn_{j}}{L_1}\right],
\quad
 U_2'=\exp\left[\i\sum_{j=1}^L\frac{2\pi jn_{j}}{L}\right].
 \label{U1'U2'def}
\end{equation}

In SBCs, the translation and polarization operators for different
directions are related as
\begin{equation}
 T_2'=T_1^{\prime L_1},\qquad U_1'=U_2^{\prime L_2}.
  \label{constraint_SBC}
\end{equation}
Then, the counterpart of Eq.(\ref{TUT.1}) becomes
\begin{equation}
 T_1' U_2' T_1^{\prime-1}
  =\e^{-\i 2\pi\hat{\rho}}U_2'.
 \label{TUT.2}
\end{equation}
We choose $U_2'$ rather than $U_1'$ here because it eliminates the
$L_2$-dependence from this relation. This was introduced for an
extension of the LSM theorem to higher dimensions\cite{Yao-O}.  The
magnetic translation operators are
\begin{equation}
 \tilde{T}_1'
=U_2^{\prime L_2\phi}T_1'
  =U_1^{\prime \phi}T_1',\qquad
  \tilde{T}_2'=\tilde{T}_1^{\prime L_1}=T_2'
  \label{mTrans_SBC}
\end{equation}
which satisfy $\mathcal{H}'=\tilde{T}_i'\mathcal{H}'\tilde{T}_i^{\prime
-1}$. Note that $L_2$-dependence in $\tilde{T}_1'$ is again removed.

The adiabatic flux insertion operator is introduced as in the same way
of Eq.~(\ref{FIb.1}) as
\begin{equation}
 F_i'(\Phi_i)=\mathcal{T}
  \exp\left[-\i\int_0^T\mathcal{H}_i'(\Phi_i\tfrac{t}{T})\d t\right]
  \label{FIa.2}
\end{equation}
and the Laughlin-type relation (\ref{FUF.1}) becomes
\begin{equation}
 [F_1'(\Phi)]^{-1}U_2'F_1'(\Phi)=U_2' \e^{+\i \nu\Phi}.
\end{equation}
In Eq.~(\ref{FIa.2}), the Hamiltonian with a flux
$\mathcal{H}_i'(\Phi_i)$ is described similarly to the case of PBC,
however, the definition Eq.~(\ref{fluxes_PBC}) for SBCs means that
effects of the fluxes $\Phi_1$ and $\Phi_2$ are no longer independent
(see also Fig.~\ref{fig:torus}(b)), because $U_1'$ and $U_2'$ have the
same phase variable except for their coefficients as in
Eq.~(\ref{U1'U2'def}). Therefore,
%
the insertion of flux $\mathcal{H}_1'(\Phi_1)$ has the same effect as
the flux $\mathcal{H}_2'(L_2\Phi_1)$ due to the relation of
Eq.~(\ref{constraint_SBC}), so that the magnetic translation operator
corresponding to Eq.~(\ref{FT.1}) is
\begin{equation}
 \bar{T}_1'=F_2'(2\pi L_2\phi)T_1'
  =F_1'(2\pi\phi)T_1'.\label{FT.2}
\end{equation}
Hence, both Eqs.~(\ref{TUT.2}) and (\ref{FT.2}) become free of $L_2$
dependence.  This is the essential advantage of SBCs.  Therefore, we
obtain the relation
\begin{equation}
 \bar{T}_1' U_2' \bar{T}_1^{\prime-1}
 =U_2'
 \e^{\i 2\pi(\phi\nu-\rho)},
\end{equation}
without any geometric parameters in contrast to Eq. (\ref{TUT.1a}). This
means that the LSM-type argument under SBCs leads to the QHE condition
$\phi\nu-\rho\in\mathbb{Z}$ {\it directly}, without passing through the
intermediate relation $L_2(\phi\nu-\rho)\in\mathbb{Z}$.

{\it --Summary and Discussion--}

In this work, we have derived the condition for the appearance of the
integer quantum Hall effect (QHE) in 2D lattice systems with arbitrary
interactions, namely $\phi \nu - \rho \in \mathbb{Z}$, by means of a
Lieb-Schultz-Mattis (LSM)-type and Laughlin-type argument.  By
introducing the \textit{spiral boundary conditions} (SBCs), we have
successfully eliminated the artificial dependence on the system width
$L_2$ that arises in the conventional argument based on periodic
boundary conditions (PBCs).
Under the SBC, the 2D lattice can be mapped onto an extended 1D chain,
so that explicit dependence on the system's cross section is avoided,
and the effects of flux insertion in different directions are treated on
equal footing.  This construction provides a more transparent physical
picture of the quantization of the Hall conductance in lattice fermion
systems.

The key to this approach is that the spatial directions of the external
force and the response can be controlled by a factor of the system size
as in Eqs.~(\ref{constraint_SBC})-(\ref{mTrans_SBC}), and (\ref{FT.2}),
thereby allowing the flux insertion process to be represented without
introducing any redundant parameters.  Consequently, the condition
$\phi\nu - \rho \in \mathbb{Z}$ emerges directly from the magnetic
translation symmetry, demonstrating that the quantization of the Hall
conductance can be understood purely from symmetry and topology, without
relying on explicit band-structure calculations.

Although we emphasize the advantages of the 1D description by SBCs, the
system remains physically equivalent to the 2D lattice, and hoppings in
the $x_2$ direction still appear as long-range hopping terms
$c^\dag_{j+L_1}\e^{\i\theta_{j}^2}c^{\mathstrut}_{j} +\mbox{H.c.}$ in
the 1D chain representation. Thus, cutting the torus under SBCs removes
not only the bond on the 1D chain, but also $L_1$ long-range bonds as
shown in Fig.~\ref{fig:unit_SBC}.  Therefore, compared with PBCs, the
number of removed bonds differs only by one, which does not affect the
physics in the thermodynamic limit. Hence, the edge-state properties
under the two boundary conditions are essentially the same, and
Laughlin's relation holds in both cases.

Our approach unifies the LSM argument for interacting systems and
Laughlin's gauge argument for the QHE, and provides a useful framework
for analyzing correlated topological phases in lattice models.  Since
the present formalism is formulated entirely in terms of the many-body
wave function and the polarization operators, it can be
straightforwardly extended to systems with disorder or nontrivial
lattice geometries, and also to the fractional QHE\cite{Lu-R-O}.
In addition to the present method for the QHE condition, an alternative
formulation has also been proposed by Matsugatani {\it et
al.}\cite{Matsugatani-I-S-W} in terms of many-body Chern number based on
Berry connections \cite{TKNN,Fukui-H-S} and point-group symmetries.  It
would be also interesting to reformulate their framework using SBCs.


{\it --Acknowledgment--}

The authors thank Y. Imamura and M. Oshikawa for discussions.
M.~N. acknowledges the research fellow position of the Institute of
Industrial Science, The University of Tokyo. M.~N. is supported partly
by MEXT/JSPS KAKENHI Grant No.~JP20K03769.

\end{document}